\documentclass[a4paper,11pt]{article}
\pdfoutput=1 

\usepackage{amsmath} 
	\allowdisplaybreaks
\usepackage{jheppub} 
\usepackage{empheq}
\usepackage{bm}
\usepackage{slashed}
\usepackage[T1]{fontenc} 
\usepackage[all]{xy}
\usepackage{rotating}
\usepackage{mathtools}
\usepackage{bbm}
\usepackage{dcolumn}
\usepackage{multirow}
\usepackage{mathrsfs}
\usepackage{arydshln}
\usepackage{verbatim}
\usepackage{graphics}
\usepackage{longtable}

\def\diag{\mathop{\rm diag}\nolimits}

\usepackage{slashed}

\def\a{\alpha}
\def\b{\beta}

\def\CC{{\cal C}}

\def\CF{{\cal F}}

\def\CH{{\cal H}}
\def\CI{{\cal I}}

\def\CM{{\cal M}}
\def\CN{{\cal N}}

\def\CS{{\cal S}}

\def\CW{{\cal W}}

\def\CZ{{\cal Z}}

\def\beq#1\eeq{\begin{align}#1\end{align}}

    \let\p=\pi

\newcommand{\be}{\begin{equation}}
	\newcommand{\ee}{\end{equation}}
\newcommand{\ba}{\begin{align}}
	\newcommand{\ea}{\end{align}}

\newcommand{\bi}{\begin{itemize}}
	\newcommand{\ei}{\end{itemize}}

\makeatletter
\newcommand*{\rom}[1]{\expandafter\romannumeral #1}
\makeatother

\title{\boldmath Generalized non-unitary Haagerup-Izumi  modular data from 3D S-fold SCFTs}

\abstract{By applying the recently proposed  (3D rank-0 $\mathcal{N}$=4 SCFT)/(non-unitary TQFTs) correspondence to  S-fold SCFTs, we construct an exotic class of non-unitary TQFTs labelled by an integer $k\geq 3$. 
	The SCFTs are obtained by gauging diagonal $SU(2)$ subgroup of $T[SU(2)]$ theory with Chern-Simons level $k$.  We give the explicit expression for modular data, $S$ and $T$ matrices, of the TQFTs. 
	When $k=4m^2+4m+3$ with an integer $m\geq 1$, the modular data (modulo a decoupled semion) is identical to  a non-unitary Haagerup-Izumi modular data. Thus, we give a physical realization of the exotic non-unitary  modular data as well as  its generalization using an exotic class of SCFTs.
	 }

\author[a]{Dongmin Gang}
\author[a]{and Dongyeob Kim}

\affiliation[a]{
	Department of Physics and Astronomy $\&$ Center for Theoretical Physics,
	\\
	Seoul National University, 1 Gwanak-ro, Seoul 08826, Korea}

\emailAdd{arima275@snu.ac.kr}
\emailAdd{ktfa159@snu.ac.kr}

\begin{document} 
	\maketitle
	\flushbottom



\section{Introduction}
Recently, there has been growing interest in   classifying conformal field theories (CFTs) and topological field theories (TQFTs) in various space-time dimensions. They describe universal macroscopic behaviors of  many-body quantum or statistical systems, ranging from critical Ising model, fractional quantum hall systems to quantum gravity via holography principle.  Exact results in TQFTs and  supersymmetric CFTs (SCFTs)  have triggered the currently ongoing explosion in the area of  physical mathematics, see \cite{Bah:2022wot} for a recent nice review on the subject.

In this paper, we focus on two corners of the big classification program,  classification of 3D $\CN=4$  rank-0 SCFTs and 3D non-unitary TQFTs, and their intricate interplay.  Being rank-0 means absence of Coulomb and Higgs branches. Despite  its simplicity,  these rank-0 theories have been overlooked in the classification program until quite recently. One reason is that most classification schemes of SCFTs with 8 supercharges rely on the existence of the vacuum moduli spaces and its geometric structures, which are absent in the rank-0 theories. Another reason is that most  rank-0 SCFTs do not allow microscopic Lagrangian description with manifest 8 supercharges. They are constructed through a non-trivial SUSY enhancement mechanism. 
 Recently, the rank-0 SCFT turns out to contain a pair of non-unitary (semi-simple) TQFTs, say ${\rm TFT}_{\pm}$ in a  Coulomb/Higgs branch limit \cite{Gang:2021hrd}.\footnote{Generally, one can consider a pair of non-unitary TQFTs associated to any $\CN=4$ SCFTs via topological twisting using $SU(2)_L$ or $SU(2)_R$ subgroup of the $SO(4) \simeq SU(2)_L\times SU(2)_R$ R-symmetry. For non-zero rank case, the resulting non-unitary TQFTs are non-semisimple and thus are not genuine TQFTs satisfying Atiyah's axioms. Understanding  mathematical structures of the generalized notion of  TQFTs and its relation to 3D SCFTs are currently active research area \cite{Gukov:2020lqm,Creutzig:2021ext,Garner:2022rwe}.  }
The (rank-0 SCFT)/(non-unitary TQFTs) correspondence provides a novel classification scheme of  rank-0 SCFTs by studying its associated non-unitary TQFTs.
 Using the correspondence, for example, one can  derive a lower bound on $F$ (round 3-sphere free-energy) for $\CN=4$ rank-0 SCFTs \cite{Gang:2021hrd}.  In this paper, 
we use the correspondence in the opposite direction and construct a new class of exotic non-unitary TQFTs from a known  class of rank-0 SCFTs called `S-fold SCFTs'.  The theory is obtained by gauging the diagonal $SU(N)$ subgroup of the $T[SU(N)]$ theory with Chern-Simons level $k$.  Various interesting aspects of the S-fold theory has been uncovered recently. For example, they appear naturally in the context of 3D-3D correspondence \cite{Terashima:2011qi},  enjoy SUSY enhancement \cite{Gang:2018huc,Assel:2018vtq,Garozzo:2019ejm}, have  non-geometrical holographic dual \cite{Assel:2018vtq} and  have exactly marginal deformation preserving $\CN=2$ superconformal algebra \cite{Beratto:2020qyk}. We focus on the case when $N=2$ and the S-fold SCFTs will be denoted by $\mathcal{S}_k$. 
 Based on  the exact computations of  BPS partition functions combined with basic dictionaries of the correspondence and consistency conditions, we  propose the explicit expression for modular data, $S$ and $T$ matrices, of the non-unitary TQFT, ${\rm TFT}[\CS_k]$ for all $k\geq 3$. The modular data is given in  \eqref{full S and T}.  When $k=4 m^2+4m+3 $ with $m\in \mathbb{N}$, interestingly, the modular data is identical to a non-unitary Haagerup-Izumi modular data \cite{haagerup1994principal,haagerup1999exotic,izumi2000structure,Evans:2010yr,Evans:2015zga} modulo a decoupled $U(1)_{2}$ or $U(1)_{-2}$.   The  modular data draws much attention  since it contains exotic fusion algebra (i.e. generalized symmetry) which can not be constructed in the conventional approaches based on finite group or affine Lie algebra.  
 We give a physical realization of the exotic modular data from a Coulomb or Higgs branch limit of the   exotic (rank-0) 3D SCFTs\footnote{Recently, a lattice model realization of  a 2D CFT with Haagerup fusion algebra  are studied \cite{Huang:2021nvb,Vanhove:2021zop}. The Haagerup-Izumi  modular data is a generalization of  quantum double of  the Haagerup fusion category.  }  and  generalizes it to arbitrary $k \geq 3$.

\section{Non-unitary TQFTs from S-fold SCFTs}

We review a field theoretic construction of S-fold SCFTs and its various supersymmetric partition functions. From the computation, we extract the partial information on the modular data  of the non-unitary TQFTs appearing in the Coulomb/Higgs branch limit. By imposing universal properties of modular data on the top of the partial information, we can determine the full modular data given in \eqref{full S and T}. 

\subsection{3D S-fold SCFTs and its BPS partition functions}
The $T[SU(2)]$ theory is a 3D $\mathcal{N}=4$ $U(1)$ gauge theory with 2 fundamental hypermultiplets \cite{Gaiotto:2008ak}.  The theory has UV $SU(2)_H\times U(1)_C$ flavor symmetry which is enhanced to $SU(2)_H\times SU(2)_C$ in IR. The theory has vacuum moduli space $\CH\times \CC$ where both of Coulomb branch $\CC$ and Higgs branch $\CH$ are $\mathbb{C}^2/\mathbb{Z}_2$. The Coulomb branch (resp. Higgs branch) is parametrized by scalar fields charged under $SU(2)_C$ (resp. $SU(2)_H$). By gauging the diagonal $SU(2)^{\rm diag}$ subgroup of the $SU(2)_H\times SU(2)_C$ with a non-zero Chern-Simons level $k$,  all the vacuum moduli are lifted. 
Generally gauging with non-zero CS level $k$ breaks $\CN=4$ supersymmetries to $\CN=3$. But thanks to the nilpotency property of the momenta maps, $\mu_H$ and $\mu_C$, of the $SU(2)_H \times SU(2)_C$ symmetry, the gauging does not break any SUSY. 
When $|k|<2$, the gauge theory  has a mass gap and flows to a unitary topological field theory (TQFT) in IR.  For $|k|=2$, there is a decoupled free hypermultiplet. 
For $|k|\geq 3$, the theory flows to a rank-0 $\CN=4$ SCFT called `S-fold SCFT'.  
\begin{align}
	\textrm{S-fold SCFT : } \CS_{k:|k|\geq 3}:=\frac{T[SU(2)]}{SU(2)^{\rm diag}_{k}}\;.
\end{align}
The  (rank-0 SCFT)/(non-unitary TQFTs) correspondence predicts there is a pair of non-unitary TQFTs, ${\rm TFT}_\pm [\CS_k]$, associated with SCFTs. Due to the self-mirror property of the $T[SU(2)]$ theory, the two topological field theories are identical and  will be denoted simply by ${\rm TFT}[\CS_k]$.   To extract the modular data of the TQFT, we compute the BPS partition functions using localization.  First, the squashed 3-sphere partition function $\CZ^{S^3_b}_{\CS_k} (m, \nu)$ is  \cite{Kapustin:2009kz,Jafferis:2010un,Hama:2010av,Hama:2011ea}
 \begin{align}
 \begin{split}
 	&\CZ^{S^3_b}_{\CS_k}(m,\nu) = \frac{1}2 \int \frac{dX dZ}{2\pi \hbar} \CI_\hbar(X,Z;W)\big{|}_{W =  m + \nu (i\pi +\frac{\hbar}2)} \;, \textrm{ where}
 	\\
 	&\CI_\hbar (X, Z;W) = 4 \sinh (X) \sinh \left(\frac{2\pi i X}{\hbar} \right)\exp \left( \frac{2k X^2 +2 (X-Z)^2+ W^2-(i\pi +\frac{\hbar}2)W}{2\hbar}\right)
 	\\
 	&\qquad \qquad  \quad\times \left( \prod_{\epsilon_1, \epsilon_2 = \pm 1} \psi_\hbar \left(\epsilon_1 Z+\epsilon_2 X+\frac{W+i \pi +\hbar/2}2 \right)  \right) \psi_\hbar (-W+i \pi +\frac{\hbar}2) \;.
 	\end{split}
 \end{align}
 Here $\psi_\hbar$ is the non-compact quantum dilogarithm (Q.D.L) function and $\hbar = 2\pi ib^2$ where $b$ is the squashing parameter.  We follow the notation as in \cite{Gang:2019jut}. The S-fold theory has $\CN=4$ superconformal symmetry which includes  $SO(4) \simeq SU(2)_L\times SU(2)_R$ R-symmetry. In terms of the $\CN=2$ subalgebra on which the localization formula is based,  the two Cartans of $SO(4)$ R-symmetry correspond to the superconformal $U(1)_R$ symmetry and a flavor symmetry $U(1)_A$ called `axial symmetry'. The charges, $R$ and $A$, of the $U(1)_R$ and $U(1)_A$ are 
 \begin{align}
 R = J_3^L+J_3^R\;, \quad A = J_3^L-J_3^R\;.
 \end{align}
Here $J_3^{L} \in \mathbb{Z}/2$ and $J_3^R \in \mathbb{Z}/2$ are the  Cartan of $SU(2)^L$ and $SU(2)^R$ respectively. In the localization computation, one needs to choose a $U(1)$ R symmetry, which is not necessarily identical  to the superconformal R-symmetry. The general choice of $U(1)$ symmetry is given by a linear combination of the $U(1)_R$ and $U(1)_A$ symmetry parametrized by a mixing parameter $\nu \in \mathbb{R}$, whose charge is given as
\begin{align}
	R_\nu = R+\nu A\;.
\end{align}
$\nu=0$ corresponds to the superconformal R-symmetry while  $R_{\nu=1} = 2J_3^L$ and $R_{\nu=-1} = 2J_3^R$. $m$ in the above is the real mass for the $U(1)_A$ symmetry.

 By expanding the integrand in the asymptotic limit $\hbar \rightarrow0 $, one obtains quantum twisted superpotential
 \begin{align}
 	\log \CI_\hbar (X,Z;W  = m + \nu (i\pi +\frac{\hbar}2))\xrightarrow{\quad \hbar\rightarrow 0 \quad } \frac{1}\hbar \CW_0 (X,Z;m, \nu)+ \CW_1 (X, Z;m,\nu)+O(\hbar)\;.
 \end{align}
For computation, use the following asymptotic expansion of the Q.D.L. 
\begin{align}
	\log \psi_\hbar (Z) \xrightarrow{\quad \hbar\rightarrow 0 \quad } \frac{1}\hbar {\rm Li}_2 (e^{-X}) - \frac{1}2 \log (1-e^{-X}) +O(\hbar)\;.
\end{align}
From the first two terms $\{\CW_0, \CW_1\}$ in the expansion, one can obtain Bethe-vacua $(z_\alpha, x_\alpha) \in S_{\rm B.E.}$ and their handle-gluing $\CH_\alpha$ and fibering $\CF_\alpha$ as follows \cite{Closset:2016arn,Closset:2017zgf,Closset:2018ghr,Gang:2019jut}
\begin{align}
\begin{split}
S_{\rm B.E.} (m,\nu)&= \big{\{} (z,x)\;:\;  \left(\exp (\partial_Z \CW_0), \exp (\partial_X \CW_0) \right)|_{Z\rightarrow \log z, X\rightarrow \log x} = (1,1), \; x^2 \neq 1 \big{\}} /\mathbb{Z}^{\rm Weyl}_2
\\
&= \{ (z_\alpha, x_\alpha)\}_{\alpha=0}^{r-1}\;,
\\
\CH_\alpha (m, \nu)& =\exp (-2 \CW_1 )\det_{i,j} \partial_{i} \partial_j \CW_0\big{|}_{Z\rightarrow \log z_\alpha, X \rightarrow \log x_\alpha}\; (\textrm{we define }\partial_1 := \partial_Z, \; \partial_2 := \partial_X)\;,
\\
\CF_\alpha (m, \nu)  & = \exp \left(-\frac{ \CW_0-2\p i \vec{n}_\alpha \cdot (Z, X)}{2\pi  i}\right)  \bigg{|}_{Z\rightarrow \log z_\alpha, X \rightarrow \log x_\alpha}\;.
\end{split}
\end{align}
In the computation of the fibering operator, the integer-valued vector $\vec{n}_\alpha = (n_\alpha^Z, n_\alpha^W)$ is defined by 
\begin{align}
 \partial_i (\CW_0 - 2\pi i \vec{n}_\alpha \cdot (Z,X))|_{Z\rightarrow \log z_\alpha, X\rightarrow \log x_\alpha} = 0\;, \quad i=1,2\;.
\end{align}
The $\mathbb{Z}_2^{\rm Weyl}$ symmetry act as 
\begin{align}
 \mathbb{Z}_2^{\rm Weyl} \;;\; x \leftrightarrow \frac{1}x\;.
\end{align}
and we need to discard the Bethe-vacuum  invariant under the symmetry, i.e. $x^2=1$.
Using them, we can compute the twisted partition function $\CZ^{\CM_{g,p}} (m, \nu)$
\begin{align}
\CZ^{\CM_{g,p}} (m, \nu) = \sum_{\alpha \in S_{\rm B.E.} (m,\nu)} \CH_\alpha(m, \nu)^{g-1} \CF_\alpha (m, \nu)^{p}\;. \label{twisted ptns}
\end{align}
The $\CM_{g,p}$ denotes the degree $p$ bundle over genus-$g$ Riemann surface $\Sigma_g$.  To preserve some supercharges, we turned on a background magnetic monopole flux coupled to the $U(1)_{R_\nu}$ symmetry. 
%
%

Using the above expressions, one can compute the $\CH_\a $ and $\CF_\alpha$ for the $\CS_k$ theory. There are $(2k+2)$ Bethe-vacua and their handle-gluing/fibering operator at $\nu=\pm 1 $ and $m=0$ are  
\begin{align}
\begin{split}
&\{\CH_\alpha (m=0, \nu  = \pm 1)\}_{\alpha = 0}^{2k+1} =  \{ (a_0^{-2})^{\otimes 2}, (a_1^{-2})^{\otimes (k-3)},(a_2^{-2})^{\otimes (k+1)},(a_3^{-2})^{\otimes 2} \}\;,
\\
& \{ \CF_\alpha (m=0, \nu= \pm 1 )\}_{\alpha=0}^{2k+1}  = \{e^{2\pi i \delta }\exp (2\pi i h_\alpha)\}_{\alpha=0}^{2k+1}\;  \textrm{ with }
\\
&\{ h_\alpha\}_{\alpha=0}^{2k+1} =\bigg{ \{} 0, \frac{k+2}{4}, \frac{A^2}{4(k-2)}\bigg{|}_{A=1,\ldots, k-3},  \frac{B^2}{4(k+2)} \bigg{|}_{B=1,\ldots, k+1},\frac{k+2}4 , 0 \bigg{\}} \;. \label{H and F}
\end{split}
\end{align}
$\delta$ is a rational number which is sensitive to local counter-terms. 
Here $a_{0,1,2,3}$ are
\begin{align}
\begin{split}
&(a_0,a_1, a_2, a_3) 
\\
&=\left(\frac{1}{\sqrt{8(k-2)}} +  \frac{1}{\sqrt{8(k+2)}}, \frac{1}{\sqrt{2(k-2)}}, \frac{1}{\sqrt{2(k+2)}}, \; \frac{1}{\sqrt{8(k-2)}} -  \frac{1}{\sqrt{8(k+2)}}\right). \label{a0a1a2a3}
\end{split}
\end{align}
\subsection{Modular data of ${\rm TFT}[\CS_{k \geq 3}]$}
As a main dictionary of the (rank-0 SCFT)/(non-unitary TQFTs) correspondence, it is proposed that
\begin{align}
\CZ^{\CM_{g,p}}_{\CS_k} (m=0, \nu= \pm 1) =   \CZ^{\CM_{g,p}}_{{\rm TFT}[\CS_k]}\;. \label{dictionary}
\end{align}
Here $\CZ^{\CM_{g,p}}_{{\rm TFT}[\CS_k]}$ is  partition function of  the non-unitary topological field theory  ${\rm TFT}[S_k]$ on $\CM_{g,p}$.  For general bosonic topological field theories, on the other hand, the partition function can be given in terms of the modular data, $S$ and $T$ matrices,  as follows
\begin{align}
\CZ^{\CM_{g,p}}_{\rm TFT} = \sum_{\alpha=0}^{r-1} (S_{0\alpha})^{2(1-g)} (T_{\alpha\alpha})^{p}\;. \label{Mgp for general TQFT}
\end{align}
Here $r$ is the rank, i.e. the dimension of Hilbert-space on two-torus,  of topological field theory.  Generally, the $S$ matrix is symmetric and unitary while the $T$ matrix  is diagonal and unitary. By combining BPS partition function results in  \eqref{twisted ptns} and \eqref{H and F} with \eqref{dictionary} and \eqref{Mgp for general TQFT}, we can extract some parts of  modular data of ${\rm TFT}[\CS_k]$ as follows
\begin{align}
\begin{split}
&\{|S_{0\alpha} (\textrm{of } {\rm TFT}[\CS_k])|\}_{\alpha=0}^{ 2k+1} = \{  a_0^{\otimes 2}, a_1^{\otimes (k-3)}, a_2^{\otimes (k+1)}, a_3^{\otimes 2}\}\;,
\\
& (T_{\alpha \beta } \textrm{ of } {\rm TFT}[\CS_k])  = \delta_{\alpha, \beta} \exp (2\pi i h_\alpha) \;, \;  \textrm{where $h_\alpha$ is given in \eqref{H and F}}\;.
\end{split}
\end{align}
A remaining non-trivial task is to find the full $S$-matrix satisfying following $SL(2,\mathbb{Z})$ relations\footnote{Conventionally, the $T$ matrix is defined as $T_{\alpha \beta} = \delta_{\alpha, \beta}\exp \left(2\pi i (h_\alpha- \frac{c_{2d}}{24}) \right)$ and satisfies $(ST)^3= C$ where $h_\alpha$ is  topological spin. Our $T$ is different from the conventional $T$ by an overall phase factor $\exp \left( \frac{2\pi i c_{2d}}{24}\right)$. }
\begin{align}
S^2 = C\;, \quad  (ST)^3 = \exp \left(\frac{2\pi i c_{2d}}{8} \right)C\;, \label{SL(2,Z) relation}
\end{align}
where $C$ is a permutation matrix called charge conjugation and $c_{2d}\in \mathbb{Q}$ is so-called 2d chiral central charge defined only modulo $8$.  A priori, there is no reason to expect that the partial modular data obtained from the handle gluing/fibering operators of a  supersymmetric gauge theory can be completed into unitary $S$ and $T$ matrices forming a $SL(2,\mathbb{Z})$ representation.  To be completed to full $SL(2,\mathbb{Z})$, for example,  the handle-gluing/fibering operators should satisfy  following non-trivial conditions 
\begin{align}
\begin{split}
	&|\CF_\alpha |=1, \quad \sum_\alpha \CH_\alpha^{-1} =1 \;,
	\\
	&\textrm{ and } \exists \textrm{  a Bethe-vacuum $\alpha=0$ satsifying } |\sum_\alpha \CH^{-1}_\alpha  \CF_\alpha| = \CH_{\alpha=0}^{-1/2}\;.
\end{split}
\end{align}
These conditions are not satisfied for generic $\CN=2$ or $\CN=4$ theories.  The first condition implies that the diagonal $T$-matrix is unitary,  2nd means $(S^2)_{00}=1$ and the last one implies that $|(STS)_{00}| =| (T^{-1}S^{-1}T^{-1})_{00}| $. All the  conditions above are  satisfied for the handle-gluing/fibering operators in \eqref{H and F}. The  (rank-0 SCFT)/(non-unitary TQFTs) correspondence further predicts that   for 3D $\CN=4$ rank-0 SCFTs the full $SL(2,\mathbb{Z})$ completion is always possible  and it can be embedded into a non-unitary modular tensor category. 

As a main result of this paper, we propose following modular data of the non-unitary TQFT ${\rm TFT}[\CS_k]$:
\begin{align}
\begin{split}
&(S  \textrm { of  }{\rm TFT}[\CS_k])
\\
&= \setlength\arraycolsep{1pt} \left(\begin{array}{c|c|c|c}
\begin{matrix}
a_0 & a_0 \\ a_0 & (-1)^k a_0
\end{matrix} & \begin{matrix}
{a_1} & a_1 & a_1 &\cdots &a_1 \\  -a_1&  a_1 & -a_1 &\cdots  &  {(-1)^{k-3} a_1} \end{matrix} & \begin{matrix}
-{a_2}  & -a_2 & -a_2 \cdots & -{a_2} \\ {a_2} & -a_2 & a_2   \cdots &  {(-1)^{k+2} a_2}
\end{matrix} & \begin{matrix}
{a_3} & {a_3} \\  (-1)^k {a_3} & {a_3}
 \end{matrix}  \\
 \hline
 \begin{matrix}
a_1 & -a_1 \\ a_1 & a_1 \\  a_1 & -a_1 \\   \vdots & \vdots \\ a_1 & (-1)^{k-3} a_1
\end{matrix} &
\mbox{\Large $2 a_1 \cos{\frac{i j \pi}{k-2} } $}\bigg{|}_{1\leq i, j\leq k-3} & \mbox{\Large 0} & 
\begin{matrix}
-a_1 & a_1  \\  a_1 & a_1  \\  -a_1 & a_1  \\ \vdots & \vdots \\ (-1)^{k-3}a_1 & a_1
 \end{matrix}  \\
 \hline
 \begin{matrix}
-a_2 & a_2 \\  -a_2 & -a_2 \\  -a_2 & a_2 \\  \vdots & \vdots  \\ -a_2 & (-1)^{k+2}a_2
\end{matrix} & 
\mbox{\Large 0} & \mbox{\Large $ 2 a_2 \cos{\frac{i j \pi}{k+2} } $} \bigg{|}_{1\leq i, j\leq k+1} & \begin{matrix}
-a_2 & a_2 \\  a_2 & a_2 \\  -a_2 & a_2 \\ \vdots & \vdots\\ (-1)^{k+1}a_2 & a_2
 \end{matrix} \\
 \hline
\begin{matrix}a_3 & (-1)^k a_3 \\ a_3 & a_3 \end{matrix} & \begin{matrix}  {-a_1} & a_1 & - a_1 & \cdots & {(-1)^{k-3}a_1} \\ {a_1} & a_1 & a_1 & \cdots & {a_1} \end{matrix} & \begin{matrix} -{a_2} & a_2 & -a_2  & \cdots & {(-1)^{k+1}a_2} \\ {a_2} & a_2 & a_2 &\cdots & {a_2} \end{matrix} & \begin{matrix} {(-1)^k a_0} & {a_0} \\ {a_0} & {a_0}\end{matrix}
\end{array}\right) , 
\\
& (T  \textrm{ of  }{\rm TFT}[\CS_k])
\\
&=\diag \left[   \exp \left(  2\pi i \bigg{ \{} 0, \frac{k+2}{4}, \frac{A^2}{4(k-2)}\bigg{|}_{A=1,\ldots, k-3},  \frac{B^2}{4(k+2)} \bigg{|}_{B=1,\ldots, k+1},\frac{k+2}4 , 0  \bigg{\}}  \right) \right] \;.
\end{split}
\label{full S and T}
\end{align} 
Here $\{a_{i}\}_{i=0}^3$ are given in  \eqref{a0a1a2a3}.
There are $2+(k-3)+(k+1)+ 2 = 2(k+1)$ simple objects which will be denoted as
\begin{align}
\{\mathbf{1}, \mathbf{1}'\}, \;  \{I_i\}_{i=1}^{k-3}, \; \{J_i\}_{i=1}^{k+1}  \textrm{ and }  \{ V', V \}\;.
\end{align}
One can check that the modular matrices satisfy the $SL(2,\mathbb{Z})$ relations in \eqref{SL(2,Z) relation} with $C=(\textrm{identity})$ and $c_{2d} =1$. The fusion coefficients $N_{\alpha \beta}^\gamma$ can be computed using the Verlinde formula
\begin{align}
N_{\alpha \beta}^\gamma = \sum_\sigma \frac{S_{\a \sigma}S_{\b \sigma}S^*_{\gamma \sigma}}{S_{0\sigma}}
\end{align}
and the non-trivial fusion coefficients are 
\begin{align}  
\begin{split}
&\textrm{For even $k$},
\\
&[\mathbf{1}']\times  [\mathbf{1}'] = [\mathbf{1}]\;, \quad [\mathbf{1}']\times [I_i] = [I_{k-2-i}]\;, \quad [\mathbf{1}']\times [J_i] =  [J_{k+2-i}]\;, \quad
\\
& [\mathbf{1}']\times [V] = [V']\;, \quad  [\mathbf{1}']\times [V'] = [V]\;,  
\\
& [V]\times [I_i] = [I_i] + \sum_{j: i+j =\textrm{even}}([I_j]+[J_j])+ \begin{cases} [V]+[V'], \quad i=\textrm{even}
\\
0\;, \quad i=\textrm{odd}\end{cases}\;,
\\
& [V]\times [J_i] =- [J_i] + \sum_{j: i+j =\textrm{even}}([I_j]+[J_j])+ \begin{cases} [V]+[V'], \quad i=\textrm{even}
\\
0\;, \quad i=\textrm{odd}\end{cases}\;,
\\
& [V']\times [I_i] = [I_{k-2-i}] + \sum_{j: i+j =\textrm{even}}([I_j]+[J_j])+ \begin{cases} [V]+[V'], \quad i=\textrm{even}
\\
0\;, \quad i=\textrm{odd}\end{cases}\;,
\\
& [V']\times [J_i] =- [J_{k+2-i}] + \sum_{j: i+j =\textrm{even}}([I_j]+[J_j])+ \begin{cases} [V]+[V'], \quad i=\textrm{even}
\\
0\;, \quad i=\textrm{odd}\end{cases}\;,
\\
& [V]\times [V] =[V']\times [V']  = [\mathbf{1}]+[V]+[V']+\sum_{i:\textrm{even}}\left( [I_i]+[J_i] \right)\;,
\\
&[V]\times [V'] = [\mathbf{1}']+[V]+[V']+ \sum_{i : \textrm{even}} \left( [I_i]+[J_i] \right)\;,
\\
&[I_i]\times [I_j] = \delta_{i,j} ([\mathbf{1}] +[V])+ \delta_{i+j,k-2}([\mathbf{1}']+[V'])+  \sum_{l \;:\; i+j+l = \textrm{even}}([I_l]+[J_l])
\\&  \qquad \qquad \quad +    \sum_{l: |j-l|=i \textrm{ or }j+l = \pm i (\textrm{mod } 2k-4)}  [I_l] +    \begin{cases}[V]+[V'],  \; i+j=\textrm{even}
	\\
	0, \;  \; i+j=\textrm{odd}
\end{cases} \;,
\\
&[J_i]\times [J_j] = \delta_{i,j} ([\mathbf{1}] -[V])+ \delta_{i+j,k+2}([\mathbf{1}']-[V'])+  \sum_{l \;:\; i+j+l = \textrm{even}}([I_l]+[J_l])
\\&  \qquad \qquad \quad -    \sum_{l: |j-l|=i \textrm{ or }j+l = \pm i (\textrm{mod } 2k+4)}  [J_l] +   \begin{cases}[V]+[V'],  \; i+j=\textrm{even}
	\\
	0, \;  \; i+j=\textrm{odd}
\end{cases} \;,
\\
&[I_i]\times [J_j] = \sum_{l\;:\; i+j+l= \textrm{even}}([I_l]+[J_l])+  \begin{cases}[V]+[V'],  \; i+j=\textrm{even}
	\\
	0, \;  \; i+j=\textrm{odd}
\end{cases} \;
\\
\\
&\textrm{For odd $k$},
\\
&[\mathbf{1}']\times  [\mathbf{1}'] = [\mathbf{1}]\;, \quad [\mathbf{1}']\times [I_i] = [I_{k-2-i}]\;, \quad [\mathbf{1}']\times [J_i] =  [J_{k+2-i}]\;, \quad
\\
& [\mathbf{1}']\times [V] = [V']\;, \quad  [\mathbf{1}']\times [V'] = [V]\;,  
\\
& [V]\times [I_i] = [I_i] + \sum_{j: i+j =\textrm{even}}([I_j]+[J_j])+ \begin{cases} [V], \quad i=\textrm{even}
\\
[V']\;, \quad i=\textrm{odd}\end{cases}\;,
\\
& [V]\times [J_i] =- [J_i] + \sum_{j: i+j =\textrm{even}}([I_j]+[J_j])+ \begin{cases} [V], \quad i=\textrm{even}
\\
[V']\;, \quad i=\textrm{odd}\end{cases}\;,
\end{split}
\end{align}
\begin{align}  
\begin{split}
& [V']\times [I_i] = [I_{k-2-i}] + \sum_{j: i+j =\textrm{odd}}([I_j]+[J_j])+ \begin{cases} [V'], \quad i=\textrm{even}
\\
[V]\;, \quad i=\textrm{odd}\end{cases}\;,
\\
& [V']\times [J_i] =- [J_{k+2-i}] + \sum_{j: i+j =\textrm{odd}}([I_j]+[J_j])+ \begin{cases} [V'], \quad i=\textrm{even}
\\
[V]\;, \quad i=\textrm{odd}\end{cases}\;,
\\
& [V]\times [V] =[V']\times [V']  = [\mathbf{1}]+[V]+\sum_{i:\textrm{even}}\left( [I_i]+[J_i] \right)\;,
\\
&[V]\times [V'] = [\mathbf{1}']+[V']+ \sum_{i : \textrm{odd}} \left( [I_i]+[J_i] \right)\;,
\\
&[I_i]\times [I_j] = \delta_{i,j} ([\mathbf{1}] +[V])+ \delta_{i+j,k-2}([\mathbf{1}']+[V'])+  \sum_{l \;:\; i+j+l = \textrm{even}}([I_l]+[J_l])
\\&  \qquad \qquad \quad +    \sum_{l: |j-l|=i \textrm{ or }j+l = \pm i (\textrm{mod } 2k-4)}  [I_l] +    \begin{cases}[V],  \; i+j=\textrm{even}
	\\
	[V'], \;  \; i+j=\textrm{odd}
\end{cases} \;,
\\
&[J_i]\times [J_j] = \delta_{i,j} ([\mathbf{1}] -[V])+ \delta_{i+j,k+2}([\mathbf{1}']-[V'])+  \sum_{l \;:\; i+j+l = \textrm{even}}([I_l]+[J_l])
\\&  \qquad \qquad \quad -    \sum_{l: |j-l|=i \textrm{ or }j+l = \pm i (\textrm{mod } 2k+4)}  [J_l] +   \begin{cases}[V],  \; i+j=\textrm{even}
	\\
	[V'], \;  \; i+j=\textrm{odd}
\end{cases} \;,
\\
&[I_i]\times [J_j] = \sum_{l\;:\; i+j+l= \textrm{even}}([I_l]+[J_l])+  \begin{cases}[V],  \; i+j=\textrm{even}
	\\
	[V'], \;  \; i+j=\textrm{odd}
\end{cases} \;.
\end{split}
\end{align}
Note that all the fusion coefficients are non-negative integers.

For lower $k$ ($3\leq k \leq 5$),  the modular data of ${\rm TFT}[\CS_k]$ can be identified with that of following known non-unitary TQFTs. 
\begin{align}
\begin{split}
&{\rm TFT} [\CS_{k=3}]= (\textrm{Lee-Yang TQFT})\otimes  \overline{(\textrm{Lee-Yang TQFT})} \otimes U(1)_2\;,
\\
&{\rm TFT} [\CS_{k=4}]=\frac{(\textrm{a Galois conjugate of $SU(2)_{10}$ }) \otimes SU(2)_2}{\mathbb{Z}_2^{\rm diag}}\;,
\\
&{\rm TFT} [\CS_{k=5}]=(\textrm{a Galois conjugate of $(G_2)_{3}$ }) \otimes U(1)_{-2}\;.
\end{split}
\end{align}

\paragraph{$\mathbb{Z}_2$ 1-form symmetry and its gauging} The ${\rm TFT}[\CS_k]$ has 1-form $\mathbb{Z}_2$ symmetry originated form the center $\mathbb{Z}_2$ of the $SU(2)$ gauge symmetry in $\CS_k$. The 1-form symmetry is generated by the anyon $I'$ with topological spin $h_{\alpha=[I']} = \frac{k+2}4$. 

For odd $k$, the 1-form symmetry is anomalous \cite{Hsin:2018vcg} and its `t Hooft anomaly is the same as in  $U(1)_2$ or $U(1)_{-2}$ theory.  Actually, the theory has a decoupled $U(1)_{\pm 2}$ factor which give the anomalous 1-form symmetry:
\begin{align}
	\textrm{TFT}[\CS_k] = \begin{cases} {\rm TFT}_0 [\CS_k] \otimes U(1)_{ 2}\;,  \quad k \in 4\mathbb{Z}-1
		\\
		{\rm TFT}_0[\CS_k ]\otimes U(1)_{-2}\;, \quad k \in 4\mathbb{Z}+1\;.
		\end{cases}
\end{align}
The modular data of the non-unitary bosonic TQFT ${\rm TFT}_0 [\CS_k]$ is 
\begin{align}
\begin{split}
	&(S \textrm{ of } {\rm TFT}_0[\CS_k])|_{k \in 2\mathbb{Z}+1} = \sqrt{2}\left(\begin{array}{c|c|c}
        \begin{matrix}
         a_0 & a_3 \\
         a_3 & a_0
        \end{matrix} &  \begin{matrix}
         a_1 & \cdots & a_1 \\
         a_1 & \cdots & a_1
        \end{matrix}   &  \begin{matrix}
         -a_2 & \cdots & -a_2 \\
         a_2 & \cdots & a_2
        \end{matrix}\\
         \hline
    \begin{matrix}
         a_1 & a_1 \\ \vdots & \vdots \\ a_1 & a_1
    \end{matrix} & \mbox{\large $ 2 a_1 \cos{\frac{4 i j \pi}{k-2} } $}\bigg{|}_{1\leq i,j \leq \frac{k-3}2}  &   \mbox{\large $ 0 $} \\
         \hline 
     \begin{matrix}
         -a_2 & a_2 \\ \vdots & \vdots \\ -a_2 & a_2
     \end{matrix} & \mbox{\large $ 0 $ } & 
     \mbox{\large $ 2 a_2 \cos{\frac{4 i j \pi}{k+2} } $}\bigg{|}_{1\leq i,j \leq \frac{k+1}2}  
    \end{array}\right),
	\\
    &(T \textrm{ of } {\rm TFT}_0[\CS_k])|_{k \in 2\mathbb{Z}+1} = \mathrm{diag}\left[\exp\left(2 \pi i \left\{0,0,\frac{A^2}{(k-2)}\bigg{|}_{A=1,\cdots,\frac{k-3}{2}}, \frac{B^2}{(k+2)}\bigg{|}_{B=1,\cdots,\frac{k+1}{2}}\right\}\right)\right].
    \end{split}
    \label{odd S and T}
\end{align}
When $k=4 n^2+4n+3 $ with $n\in \mathbb{N}$,  the modular data is identical to a non-unitary Haagerup-Izumi modular data \cite{haagerup1994principal,haagerup1999exotic,izumi2000structure,Evans:2010yr,Evans:2015zga}. 

In the case $k = 4 n^2 + 4 n + 3 = (2 n + 1)^2 + 2$, the topological spin for $n$ anyons corresponding to $A = i(2 n + 1), \; i = 1, \cdots, n$ is $0\mod 1$. Thus, if we rearrange indices so that $T$ matrix becomes $\diag [1,1;\underbrace{1,1,\cdots, 1}_n, \cdots ]$, $S$-matrix becomes
\begin{align}
S = \sqrt{2} \left(\begin{array}{c|c|c|c}
     \begin{matrix}
         a_0 & a_3 \\ a_3 & a_0
     \end{matrix} & a_1 \mathbf{1}_{2 \times n} & a_1 \mathbf{1}_{2 \times n(2n+1)} & \begin{matrix}
         -a_2 & \cdots & -a_2 \\ a_2 & \cdots & a_2
     \end{matrix} \\ \hline
 a_1 \mathbf{1}_{n \times 2} & a_1 \mathbf{2}_{n \times n} & D' & \mathbf{0}_{n \times m} \\ \hline 
 a_1 \mathbf{1}_{n(2n+1) \times 2} & D'^T & E' & \mathbf{0}_{n (2n+1) \times m} \\ \hline
 \begin{matrix}
     - a_2 & a_2 \\ \vdots & \vdots \\ -a_2 & a_2
 \end{matrix} & \mathbf{0}_{m \times n} & \mathbf{0}_{m \times n (2n+1)} & \mbox{\large $ 2 a_2 \cos{\frac{4 i j \pi}{k+2} } $}\bigg{|}_{1\leq i,j \leq \frac{k+1}2}
\end{array}\right)
\end{align}
for some appropriate matrices $D'$, $E'$, and $m = 2n^2 + 2n + 2$. We can see that it is identical to (6.24) of \cite{Evans:2015zga}, and is a Galois conjugate of (3.3) in \cite{Evans:2010yr} with $\omega$ = 2. The specific form of Galois conjugation \cite{Ng:2012ty, Harvey:2018rdc} is given by
\begin{align}
\begin{split}
    &\left( S \text{ of $\mathcal{D}^2 \text{Hg}_{2n+1}$ in \cite{Evans:2010yr}}\right) \cong T^{\bar{p}} S^{-1} T^p S T^{\bar{p}} S^2, \\
    &\left( T \text{ of $\mathcal{D}^2 \text{Hg}_{2n+1}$ in \cite{Evans:2010yr}}\right) \cong T^{\bar{p}},
\end{split}
\end{align}
up to permutations, for $p = 2$ and $\bar{p} = \frac{k^2-3}{2}$.

As a representation of $SL(2,\mathbb{Z}) = \langle \mathfrak{s}, \mathfrak{t}\;:\;\mathfrak{s}^4=1, \; (\mathfrak{s}\mathfrak{t})^3=\mathfrak{s}^2\rangle$, the modular data is equivalent to a direct sum of two irreducible representations $\rho_{\frac{k-1}2} \oplus  \rho_{\frac{k+3}2}$ where
\begin{align}
\begin{split}
&\rho_d (\mathfrak{s}) = \frac{1}{\sqrt{2d-1}}  \left(\begin{array}{c|c}
	\begin{matrix}
		1
	\end{matrix} &  \begin{matrix}
		\sqrt{2}   & \cdots  \cdots \qquad  &   \sqrt{2} \\
	\end{matrix}  
   \\
	\hline
	\begin{matrix}
		\sqrt{2} \\ \vdots  \\ \sqrt{2}
	\end{matrix} & \mbox{\large $ 2  \cos{\frac{4 i j \pi}{2d-1} } $}\bigg{|}_{1\leq i,j \leq d-1}   
\end{array}\right), 
\\
 & \rho_d (\mathfrak{t})= \mathrm{diag}\left[\exp\left(2 \pi i \left\{0,\frac{A^2}{2d-1}\bigg{|}_{A=1,\cdots,d-1}\right\}\right)\right].
 \end{split}
\end{align}
In \cite{ng2022reconstruction}, they study conditions for  a direct sum of irreducible $SL(2,\mathbb{Z})$ representations to be a modular data after an appropriate unitary similar transformation. The above modular data satisfies the conditions. 

For even $k$, the 1-form symmetry is non-anomalous and we consider the theory $\textrm{TFT}[\CS_k]/\mathbb{Z}_2$ after gauging the 1-form symmetry. For $k\in 4\mathbb{Z}$, the 1-form symmetry is fermionic, i.e. $h_{\alpha=[I']} = \frac{1}2 (\textrm{mod 1})$, and thus we expect the resulting TQFT are fermionic. The fermionic TQFT with $k=8$ has the same modular data as one of  newly found rank 10 TQFTs  in  \cite{Cho:2022kzf}.  For $k\in 4\mathbb{Z}+2$, the 1-form symmetry is bosonic and thus  $\textrm{TFT}[\CS_k]/\mathbb{Z}_2$  is a bosonic non-unitary TQFT and we can consider its modular data.
 The modular data after the $\mathbb{Z}_2$ gauging is   \cite{moore1989taming,Hsin:2018vcg,Delmastro:2021xox}
\begin{align}
	\begin{split}
		&(S \textrm{ of } {\rm TFT}[\CS_k]/\mathbb{Z}_2)|_{k \in 4\mathbb{Z}+2} = \\
&\setlength\arraycolsep{1pt}\left(\begin{array}{c|c|c|c|c}
\begin{matrix}
2 a_0 & 2 a_3 \\ 2 a_3 & 2 a_0
\end{matrix}   & \begin{matrix}
\phantom{11}2 a_1\phantom{11}& \cdots & \phantom{11}2 a_1\phantom{11} \\ 2 a_1 & \cdots & 2 a_1
\end{matrix} & \begin{matrix}
\phantom{1}a_1\phantom{1} & \phantom{1}a_1\phantom{1} \\ a_1 & a_1
\end{matrix} & \begin{matrix}
\phantom{1}- 2 a_2\phantom{1}& \cdots & \phantom{1}- 2 a_2\phantom{1} \\ 2 a_2 & \cdots & 2 a_2
\end{matrix} & \begin{matrix}
- a_2 & -a_2 \\ 
a_2 & a_2
\end{matrix} \\ 
\hline
\begin{matrix}
2 a_1 & 2 a_1 \\ \vdots & \vdots \\ 2 a_1 & 2 a_1
\end{matrix} & \mbox{\large $ 4 a_1 \cos \frac{i j \pi}{n}$}\bigg{|}_{1\leq i,j \leq n-1} & \begin{matrix}
2 a_1 \mathbb{J}_{n-1}  & 2 a_1 \mathbb{J}_{n-1} \end{matrix} & \mbox{\large $ 0 $} & \mbox{\large $ 0 $}\\
\hline
\begin{matrix}
a_1 & a_1 \\ a_1 & a_1
\end{matrix} & \begin{matrix}
2a_1 \mathbb{J}^T_{n-1} \\ 2a_1 \mathbb{J}^T_{n-1}
\end{matrix} & \begin{matrix}
\phantom{1}b_{1,+}\phantom{1} &\phantom{1}b_{1,-}\phantom{1} \\ b_{1,-} & b_{1,+}
\end{matrix} & \mbox{\large $ 0 $} & \begin{matrix}
\frac{i^n}{2 \sqrt{2}} & -\frac{i^n}{2 \sqrt{2}} \\ -\frac{i^n}{2 \sqrt{2}} & \frac{i^n}{2 \sqrt{2}}
\end{matrix} \\
\hline
\begin{matrix}
-2 a_2 & 2 a_2 \\ \vdots & \vdots \\ -2a_2 & 2 a_2
\end{matrix} & \mbox{\large $ 0 $} & \mbox{\large $ 0 $} & \mbox{\large $ 4 a_2 \cos \frac{ij\pi}{n+1}$}\bigg{|}_{1\leq i,j \leq n} & \begin{matrix}
2a_2 \mathbb{J}_{n} & 2a_2 \mathbb{J}_{n} \end{matrix}\\
\hline
\begin{matrix}
- a_2 & a_2 \\ -a_2 & a_2
\end{matrix} & \mbox{\large $ 0 $} & \begin{matrix}
\frac{i^n}{2 \sqrt{2}} & -\frac{i^n}{2 \sqrt{2}} \\ -\frac{i^n}{2 \sqrt{2}} & \frac{i^n}{2 \sqrt{2}}
\end{matrix} & \begin{matrix}
2a_2 \mathbb{J}^T_{n} \\ 2a_2 \mathbb{J}^T_{n}
\end{matrix} & \begin{matrix}
\phantom{1}b_{2,+}\phantom{1} & \phantom{1}b_{2,-}\phantom{1}\\ b_{2,-} & b_{2,+}
\end{matrix}
\end{array}\right),
		\\
		&(T \textrm{ of } {\rm TFT}[\CS_k]/\mathbb{Z}_2)|_{k \in 4\mathbb{Z}+2} =\\ &\phantom{T of TFT[\CS_k]/Z_2|}\mathrm{diag}\left[\exp\left(2 \pi i \left\{0,0,\frac{A^2}{(4n)}\bigg{|}_{A=1,\cdots,n-1},\frac{n}{4}^{\otimes 2}, \frac{B^2}{4(n+1)}\bigg{|}_{B=1,\cdots,n}, \frac{n+1}{4}^{\otimes 2}\right\}\right)\right],
	\end{split}
\end{align}
where
\begin{align}
    \begin{split}
        & n = \frac{k-2}{4}, \quad
        \mathbb{J}^T_n = \left( -1, 1, -1, \cdots, (-1)^n \right),
        \\ 
        &b_{1, \pm} = (-1)^n a_1 \pm \frac{i^n}{2 \sqrt{2}}, \; \textrm{ and } \;b_{2,\pm} = (-1)^{n+1} a_2 \pm \frac{(-i)^n}{2 \sqrt{2}}.
    \end{split}
\end{align}


\section{Discussion and Future directions}
In the paper, we find a novel class of exotic non-unitary TQFTs arising  in a Coulomb or Higgs limit of S-fold SCFTs with $SU(2)$ type.  There are many interesting  generalizations and future applications.

\paragraph{Generalization to $T[G]/G^{\rm diag}_k$ for general $G=A,D,E$ and holographic dual} By generalizing the work to more general S-fold SCFTs based on $A,D,E$ Lie algebra,  one can  consider exotic non-unitary TQFTs labeled by gauge group $G=A,D,E$ and CS level $k$.  The gravity dual of the S-fold SCFT for $G=A_{N-1}$ is studied in \cite{Assel:2018vtq}. It would be interesting to see the exotic fusion algebra from the holographic dual. 

\paragraph{Class R construction and its generalizations } The S-fold SCFTs have natural realization in 3D-3D correspondence \cite{Terashima:2011qi,Dimofte:2011ju,Gang:2018wek}. They correspond to  a 3-manifold,  mapping torus whose fiber is once-punctured torus \cite{Terashima:2011qi,Gang:2015wya}. By generalizing the 3-manifolds to more general mapping torus and more general type of puncture, we expect interesting rank-0 SCFTs and their associated exotic non-unitary  TQFTs can be constructed. The modular data of these non-unitary TQFTs can be obtained from  topological invariants of the 3-manifolds as studied in \cite{Cho:2020ljj}.

\paragraph{2D non-unitary RCFTs at the edge} From the bulk-edge correspondence, we expect there is exotic class of non-unitary 2D rational conformal field theories associated to the 3D non-unitary TQFTs ${\rm TFT}[\CS_k]$. It would be interesting to bootstrap the exotic RCFTs using the modular data proposed in this paper as an input. 

\section*{Acknowledgements}
We would like to thank  Hee-Cheol Kim, Sungjay Lee and Xiao-Gang Wen for useful discussion.
The work was presented at   "East Asia Joint Workshop on Field and Strings 2022". We thank the organizers and  participants.
The work of DG is supported in part by the National Research Foundation of Korea grant NRF-2021R1G1A1095318 and NRF-2022R1C1C1011979. DG also acknowledges support by Creative-Pioneering Researchers Program through Seoul National University.

\bibliographystyle{ytphys}
\bibliography{ref}
\end{document}